\documentclass[12pt]{article}
\usepackage{amsmath}
\usepackage{amssymb}

\usepackage{scicite}

\usepackage{CJK}
\usepackage{subfigure}
\usepackage{times}
\usepackage{epsfig}
\usepackage{url}

\newenvironment{sciabstract}{%
\begin{quote} \bf}
{\end{quote}}

\newcounter{lastnote}

\begin{document}


\title{ Homophyly Networks\\ -- A Structural Theory of Networks}

\author{ Angsheng Li$^{1}$,  Jiankou Li$^{1,2 }$, Yicheng Pan$^{1,3}$ \\
\normalsize{$^{1}$State Key Laboratory of Computer Science}\\
\normalsize{ Institute of Software, Chinese Academy of Sciences}\\
\normalsize{$^{2}$University of Chinese Academy of Sciences,
P. R. China}\\
\normalsize{$^{3}$State Key Laboratory of Information Security}\\
\normalsize{ Institute of Information Engineering, Chinese Academy of Sciences,
P. R. China} }


\date{}



\baselineskip24pt

\maketitle

\begin{sciabstract}

 A grand challenge in network science is apparently the missing
of a structural theory of networks. The authors have showed that the existence of community structures
is a universal phenomenon in real networks, and that neither randomness nor preferential
attachment is a mechanism of community structures of network
\footnote{ A. Li, J. Li, and Y. Pan, Community structures are definable in networks, and universal in the real world, To appear.}. This poses a fundamental question: What are the mechanisms of community
structures of real networks? Here we found that homophyly is the
mechanism of community structures and a structural theory of networks.
We proposed a homophyly model.
It was shown that networks of our model satisfy a series of new topological, probabilistic and
combinatorial principles, including a fundamental principle, a community structure principle, a degree priority principle, a widths principle, an
inclusion and infection principle, a king node principle, and a predicting principle etc, leading to
a structural theory of networks. Our model demonstrates that
homophyly is the underlying mechanism of community structures of networks, that
nodes of the same community share common
features, that power law and small world property are never
obstacles of the existence of community structures in networks, and that community structures are definable in networks.

\end{sciabstract}

The missing of a structural theory of networks hinders us from rigorous analysis of networks and networking data.
Indeed, the current tools for networking data are mainly probabilistic or statistical methods, which
apparently neglect the structures of data. However, structures are essential. In nature and society, we
observe that mechanisms determine the structures, and that structures determine the
properties, which could be a new hypothesis of the current highly connected world. Our homophyly model explores that homophyly is the
mechanism of community structures of networks, allowing us to develop a
homophyly theory of networks. The new principles such as the
fundamental principle, the
community structure principle,
the degree priority principle, the widths principle, the inclusion and infection principle, the king node principle and the predicting principle etc
we found here provide a firm first step
for a structural theory of networks that is essential to resolving new issues of networks
such as robustness, security, stability, evolutionary games, predicting and controlling of networks.

Network  has become a universal topology in science, industry,
nature and society \cite{Bar2009}. Most real networks follow a power
law degree distribution~\cite{Bar1999, Bar2009}, and satisfy a small
world phenomenon~\cite{M1967, W1998, K2000}.

Community finding has been a powerful tool for understanding the
structures of networks and has been extensively studied
~\cite{EK2010, CR2010, CA2005, RCCLP2004, CNM2004, NEW2004,
For2010}. Newman and Girvan \cite{NG2004} defined the notion of
modularity to quantitatively measure the quality of community
structure of a network. It is built based on the assumptions that
random graphs are not expected to have community structure and that
a network has a community structure, if it is far from random
graphs.

The authors proposed the notions of entropy
community structure ratio and conductance community structure ratio
of networks. In the same paper, it was verified by experiments that
the three definitions of modularity-, entropy- and
conductance-community structures are equivalent in defining
community structures in networks, that nontrivial networks of the ER
model \cite{ER1960} and the PA model \cite{Bar1999} fail to have a
community structure, and that the existence of community structures
is a universal phenomenon in real networks. This progress posed a
fundamental question: What are the mechanisms of community
structures in real networks? What structural theories of networks we can develop?

Here we found that homophyly is the natural mechanism of community
structures of real networks. We proposed a new model of networks,
the homophyly model below, by natural mechanisms of homophyly and
preferential attachment. We show that homophyly networks provide a
foundation for a new theory of networks, the local theory of
networks.

{\bf Homophyly Model}

 Real networks are from a
wide range of disciplines of both social and physical sciences. This
hints that community structures of real networks may be the result
of natural mechanisms of evolutions of networking systems in nature
and society. Therefore mechanisms of community structures of real
networks must be natural mechanisms in nature and society.

In both nature and society, whenever an individual is born, it will
be different from all the existing individuals, it may have its own
characteristics from the very beginning of its birth. An individual
with different characteristics may develop links to existing
individuals by different mechanisms, for instance, preferential
attachment or homophyly.

We propose our homophyly model based on the above intuition. It
constructs a network dynamically by steps as follows.

{\it Homophyly model} Let $a$ be a homophyly exponent, and $d$ be a
natural number.

\begin{enumerate}
\item Let $G_d$ be an initial $d$-regular graph in which each node is
associated with a distinct color, and is called seed.

For $i>d$, let $G_{i-1}$ be the graph constructed at the end of step
$i-1$, and let $p_i=\frac{1}{(\log i)^a}$.

\item At step $i$, we create a new node $v$.

\item (Preferential attachment) With probability $p_i$, $v$ chooses a new color, in which case,

\begin{enumerate}
\item we call $v$ a seed, and

\item create $d$ edges from $v$ to nodes in $G_{i-1}$ chosen with
probability proportional to the degrees in $G_{i-1}$.
\end{enumerate}

\item (Homophyly) Otherwise, then $v$ chooses an old color, in which case,

\begin{enumerate}
\item $v$ chooses randomly and uniformly an old color as its own
color, and

 \item create $d$ edges from $v$ to nodes of the same color in
$G_{i-1}$ chosen with probability proportional to the degrees in
$G_{i-1}$.

\end{enumerate}
\end{enumerate}

The homophyly model constructs networks dynamically with both
homophyly and preferential attachment as its mechanisms. It better
reflects the evolution of networking systems in nature and society.
We call the networks constructed from the homophyly model homophyly
networks.

{\bf Homophyly Theory of Networks}

\begin{figure}
  \centering
\includegraphics[width=4in]{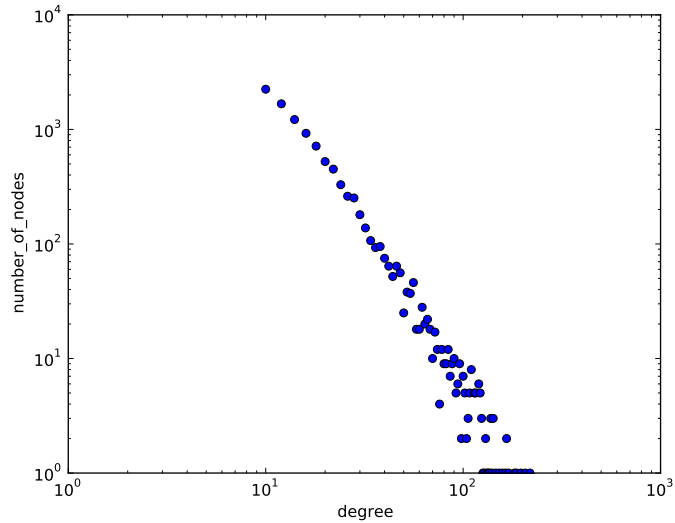}
\caption{Power law distribution of a homophyly network: $n=10,000$,
$a=1.2$ and $d=5$. }
 \label{figure_homophyly_deg_seq}
\end{figure}

\begin{figure}
  \centering
\includegraphics[width=4in]{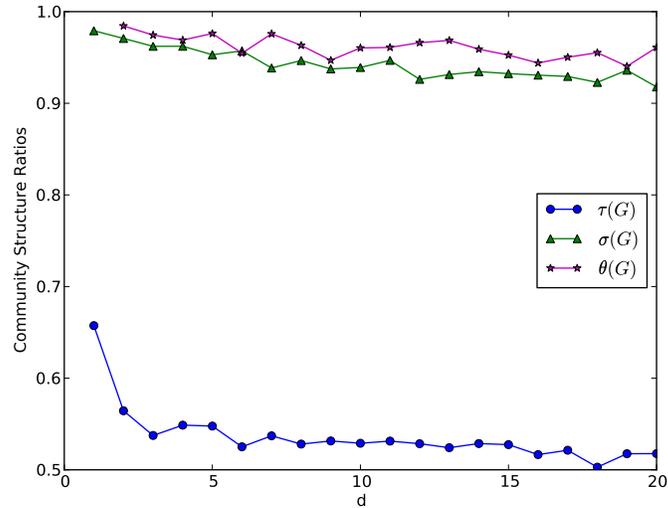}
\caption{The E-, M- and C-community structure ratios (denoted by e-,
m- and c-ratios respectively) of a homophyly network for $n=10,000$
and $a=1.2$.} \label{figure_homophyly}
\end{figure}

We will show that homophyly networks satisfy a series of new
principles, including the well known small world and power law properties.
At first, it is easy to see that the homophyly networks have
the small diameter property, which basically follows from the
classic PA model. Secondly, the networks follow a power law, for
which we see Figure \ref{figure_homophyly_deg_seq} for the intuition. At last,
they have a nice community structure, for which we depict the entropy-,  conductance-community structure ratios, and the modularity- \cite{NG2004} of some homophyly networks in Figure \ref{figure_homophyly}.
From Figure \ref{figure_homophyly}, we know that
the entropy-, modularity- and conductance-community structure ratios
of the homophyly networks are greater than $0.5$, $0.9$ and $0.9$
respectively.

Here we verify that homophyly networks satisfy a number of new topological, probabilistic
and combinatorial principles, including the fundamental principle, the community structure principle, the degree
 priority principle, the widths principle, the inclusion and infection principle, the king
node principle and the predicting principle below. (Full proofs of the principles will be referred to supplementary materials of the paper.)
We sketch the principles and their roles in network science and potential new applications as follows.

The first is a {\it fundamental principle}: Let $G=(V,E)$ be a homophyly network. Then with probability $1-o(1)$, the following properties hold:

\begin{enumerate}

\item [(1)] The number of seed nodes of $G$ is $\Omega (\frac{n}{\log^an})$.

\item [(2)] ( The small community
phenomenon~\cite{li2011community,li2012small}) The size of a
homochromatic set of $G$ is bounded by $\ln ^{\gamma}n$ for some constant $\gamma$.

 \item [(3)]  ( Power law~\cite{Bar1999}) The whole network $G$ follows a power law
 degree distribution.

\item [(4)] (Holographic law) The induced subgraph of a homochromatic set follows a
power law with the same power exponent as that of the whole network
$G$.

\item [(5)] The degrees of nodes of a homochromatic set follow a
power law.

\item [(6)] (Local communication law) The diameter of the induced subgraph
 of a homochromatic set is bounded by $O(\log\log n)$.

\item [(7)] ( The small world phenomenon~\cite{M1967, W1998, K2000}) The diameter of $G$ is bounded by
$O(\log^2 n)$.

\end{enumerate}

(1) gives an estimation of number of hubs or strong nodes in a network.
(2) shows that a community can be interpreted by the common
features of nodes in the community, that the interpretable
communities are small. (3) - (5) show that $G$ satisfy a holographic property in the sense that
the exponent of the power law of a community is the same as that of
the whole network, and that a community has
a few nodes which dominate the internal links within the community,
giving rise to an internal centrality of the communities of a
homophyly network. By (4), we may estimate the power exponent of the network by
 computing the power exponent of a community. (6) - (7) demonstrate that $G$ has the small
  world property, and that local communications within a community
are exponentially shorter than that of the global communications in the whole network $G$.
By (6) and (7), we can estimate the diameter of the network by computing the diameter of a community.

Secondly, we have a {\it community structure principle}: Let $G=(V,E)$ be a homophyly network. Then with probability $1-o(1)$,
the following properties hold:

\begin{enumerate}

\item [(1)] (Homophyly law) Let $X$ be a homochromatic set. Then the induced subgraph $G_X$ of
$X$ is connected, and the conductance of $X$, $\Phi (X)$ is bounded
by $O(\frac{1}{|X|^{\beta}})$ for some constant $\beta$.

\item [(2)] (Modularity property) The modularity \cite{NG2004} of $G$ is
$\sigma (G)=1-o(1)$.

\item [(3)] (Entropy community structure property) The entropy
community structure ratio of $G$, is $\tau
(G)=1-o(1)$.

\item [(4)] (Conductance community structure property) The conductance
community structure ratio of $G$, is $\theta
(G)=1-o(1)$.

\end{enumerate}

(1) means that a set of nodes $X$ forms a natural community if the nodes in the set share the same color, and that the conductance of a community $X$
is bounded by a number proportional to $|X|^{-\beta}$ for some constant $\beta$.
(2) - (4) show that the definitions of modularity-, entropy- and conductance-
community structure are equivalent in defining community structures in networks, and that community structures are
definable in networks.

The fundamental and community structure principles explore some basic laws governing both the global and local structures of a network.
However, to understand the roles of community structures in networks, we need to know the properties which hold for all the communities of a network.
We will see that the homophyly networks satisfy a number of such principles.

Our third principle consists of a number of properties of degrees of the networks.
Given a node $v\in V$, we define the length of degrees of $v$ to be
the number of colors associated with all the neighbors of $v$,
written by $l(v)$. For $j\leq l(v)$, we define the $j$-th degree of
$v$ to be the $j$-th largest number of edges of the form $(v,u)$'s such that the $u$'s here
share the same color, denoted by $d_j(v)$. Define the degree of $v$,
$d(v)$, to be the number of edges incident to node $v$.

Then we have a {\it degree priority principle}: Let $G=(V,E)$ be a homophyly network. Then with probability $1-o(1)$, the
degree priority of nodes in $V$ satisfies the following properties:

\begin{enumerate}

\item [(1)] (First degree property) The first degree of $v$,
$d_1(v)$ is the number of edges from $v$ to nodes of the same color
as $v$.

\item [(2)] (Second degree property) The second degree of $v$ is
bounded by a constant, i.e., $d_2(v)\leq O(1)$

\item [(3)] (The length of degrees)
\begin{enumerate}

\item The length of degrees of $v$ is bounded by $O(\log n)$.

\item Let $N$ be the number of seed nodes in $G$. For
$r=\frac{N}{\log^cN}$ for some constant $c$.
Let $x$ be a node created after time step $r$. Then the length
of degrees of $x$ is bounded by $O(\log\log n)$.

\end{enumerate}

\item [(4)] If $v$ is a seed node, then the first degree of $v$,
$d_1(v)$ is at least $\Omega (\log^{\gamma}n)$ for some constant
$\gamma$.

\end{enumerate}

The degree priority principle shows that a node $v$ has a degree priority $(d_1(v),
d_2(v),\cdots, d_l(v))$ satisfying a number of combinatorial properties, so that combinatorics
has been introduced in network theory.

Our fourth principle determines the ways of connections from a community to outside of the community.
Let $X$ be a homochromatic set of $G$. Define the {\it width of $X$
in $G$} to be the number of nodes $x$'s such that $x\in X$ and
$l(x)>1$. We use $w^G(X)$ to denote the width of $X$ in $G$.
Then we have a {\it widths Principle}: Let $G=(V,E)$ be a homophyly network. Then with probability $1-o(1)$, the following
properties hold:

\begin{enumerate}

\item [(1)] For a randomly chosen $X$, the width of $X$ in $G$ is
$w^G(X)=O(\log n)$.

Let $N$ be the number of seed nodes in $G$. For $l=N^{1-\theta}$ and
$r=\frac{N}{\log^cN}$ for some constants $\theta$ and $c$. We say
that a community is created at time step $t$, if the seed node of
the community is created at time step $t$.

\item [(2)] Let $X$ be a community created before time step $l$. Then the
width of $X$ in $G$ is at least $\Omega (\log n)$.

\item [(3)] Let $Y$ be a community created before time step $r$. Then the
width of $Y$ in $G$ is at least $\Omega (\log\log n)$

\item [(4)] Let $Z$ be a community created after time step $r$. Then the
width of $Z$ in $G$ is at most $O(\log\log n)$.

\end{enumerate}

The width of a community $X$ determines the patterns of links from
nodes in the community to nodes outside of the community.
By (4), we have that almost all communities have widths bounded by $O(\log\log n)$.
This property, together with the holographic law in the fundamental principle show that almost surely, a community has
both an internal and an external centrality. This helps us to
analyze the communications among different communities.

Our fifth principle is an inclusion and infection among the nodes of a homophyly network.
Given a node $x$ of some community $X$. We define the width of
$x$ in $G$, denoted by $w^G(x)$, is the number of communities $Y$'s such that $X\not=Y$ and such that there
is a non-seed node $y\in Y$ with which there is an edge between $x$
and $y$. Then we have an {\it inclusion and infection principle}: Let $G=(V,E)$ be a homophyly network. Then for following properties hold:

\begin{enumerate}
\item [(1)] (Inclusion property) For a non-seed node $x$ in $G$, the width of $x$ in $G$ is $w^G(x)=0$.

\item [(2)] (Widths of seed nodes) For every seed node $x$ in $G$, the width of $x$ is
bounded by $O(1)$.

\end{enumerate}

Intuitively speaking, non-seed nodes of a network are vulnerable against
attacks. In the cascading failure model of attacks, it is possible that a few number of attacks
may generate a global failure of the network. For this, one of the reasons is that the huge number of vulnerable nodes form
a giant connected component of the network, in which the attack of a few vulnerable nodes mat infect the giant connected component of
the vulnerable nodes. (1) ensures that this is not going to happen in homophyly networks.
We interpret seed nodes as strong against attacks. Let $x$ be a seed node. If $w^G(x)>1$, then it is possible for $x$ to infect two vulnerable nodes, $y_1$ and $y_2$ say, of two different communities
$Y_1$ and $Y_2$ respectively. In this case, it is easy for $y_1$ and $y_2$ to infect the seed nodes of $Y_1$ and $Y_2$ respectively. By this way, the infections of communities intrigued by the seed node $x$ may
grow exponentially in a tree of communities. (2) ensures that for each seed node $x$ of $G$, $w^G(x)=O(1)$, which is probably larger than $1$. By this reason, we know that homophyly networks are
insecure against attacks in the cascading failure models. This suggests that to make a network $G$ secure, we have to make sure that for each hub, $x$ say, the width of $x$ in $G$ is at most $1$.

Our sixth principle is the remarkable role of seed nodes in the corresponding communities and in the whole network. We have a
{\it king node principle}: Let $G=(V,E)$ be a homophyly network. Then with probability $1-o(1)$, for a community $X$ and its seed node $x_0$,
the expectation of the degree of $x_0$ is at least twice of that of the second largest degree node $x\in X$.

This principle ensures that there is a significant fraction of communities, each of which contains a king node whose degree is at least twice of that of the second largest degree node within the community.
This is a phenomenon  similar to that in a community of honey bees. It implies that
in evolutionary prisoner's dilemma games in a network, the strategies of nodes within a community could follow that of the king node,
similarly to the behaviors of a community of honey bees in nature.

The six principles above explore the mathematical properties of the
homophyly networks. They show that the community structures and properties of the communities do play essential roles
in fundamental issues and applications of networks.

Our model demonstrates that dynamic and scale-free
networks may have a community structure for which homophyly and
preferential attachment are the underlying mechanisms. This explains
the reason why most real networks have community structures and
simultaneously follow a power law and have a small world property.

{\bf Homophyly Law of Networks}

The essence of the homophyly model is the principle that: Nodes of
the same community share common features. We will show that this property provides the principle for
predicting in networks.

To verify the homophyly law, we implement an experiment of keywords
prediction in a citation network, the Arxiv HEP-TH (high energy
physics theory) citation network, which covers all the citations
within a dataset of $27,770$ papers with $352,807$ edges, in which
there are $1214$ papers that have known keywords listed by their
authors. We call the $1214$ papers annotated, and the others
un-annotated.

 We use the keywords of a paper to interpret the functions of
the paper. We predict and confirm keywords for the un-annotated
papers based on the keywords of the $1214$ annotated papers.

Let $C$ be a community found by an algorithm. For some small
constant $k$, we use the most popular $k$ keywords appeared in the
annotated papers in $C$ to represent the common features of $C$,
written ${\rm CF}(C)$. Then we predict that each keyword in ${\rm
CF}(C)$ is a keyword of an un-annotated paper in $C$.

For a keyword $K\in {\rm CF}(C)$, and a paper $P\in C$, we say that
 $K$ is confirmed to be a keyword of $P$, if $K$ appears in either
the title or the abstract of paper $P$.

\begin{figure}
\centering
\includegraphics[width=3in]{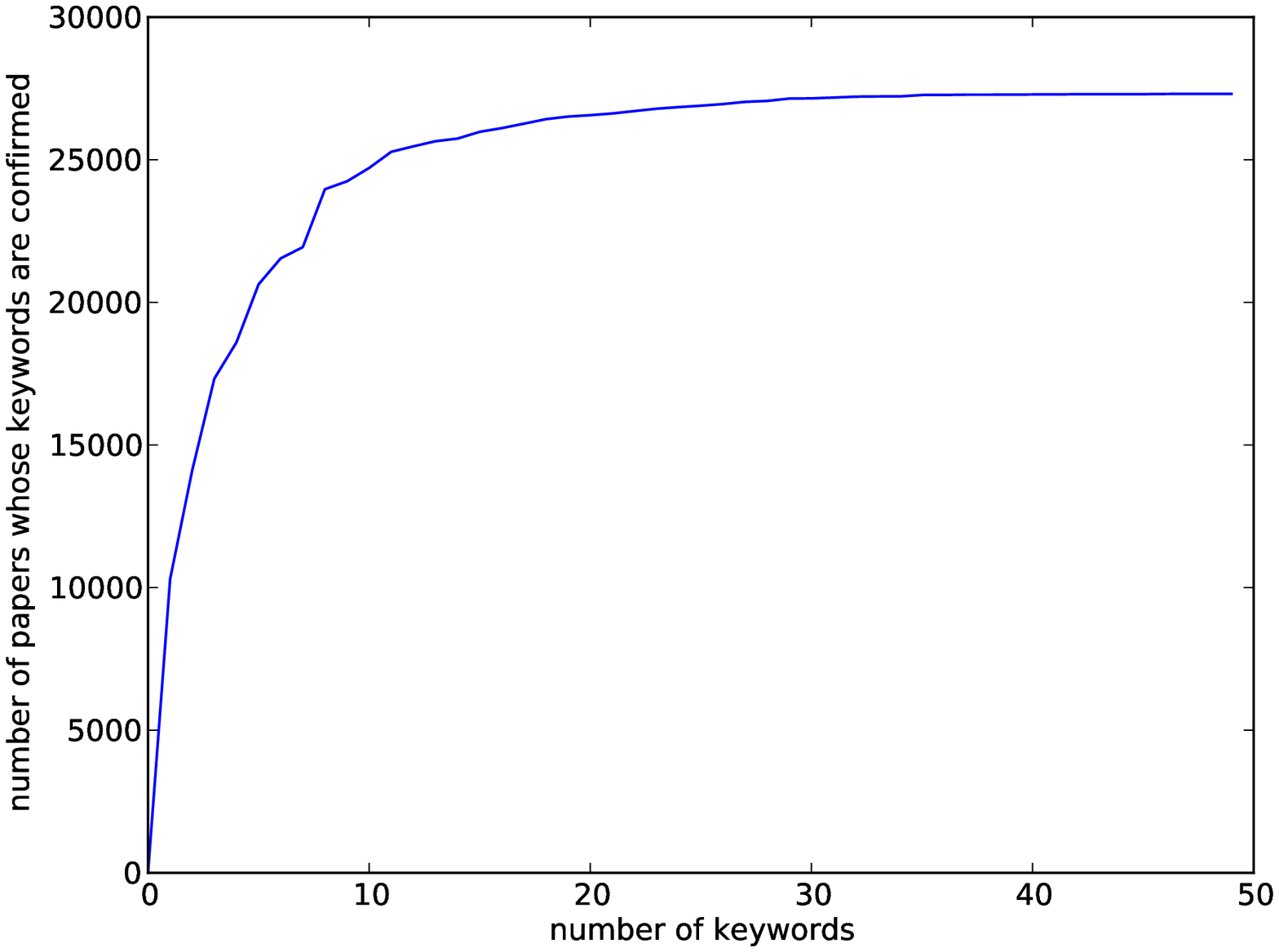}
\caption{Keywords prediction. The curve represents the numbers of
papers whose keywords are predicted and confirmed by using the most
popular $k$ keywords as the common keywords of all the communities,
for $k\leq 50$. The curve increases quickly and becomes flatten
after $k>10$. This means that each community has a few ($10$)
remarkable common keywords, a result predicted by the homophyly
networks.} \label{figure_keywords prediction 1}
\end{figure}

For each community, we use the most popular $k$ keywords appeared in
$C$ to denote the CF of the community. The full prediction and
confirmation of keywords
  by taking the most popular $i$ keywords as the CF for each community, for all
  possible $i$, is depicted in Figure~\ref{figure_keywords prediction 1}. From the figure, we
observe that for each of the communities, we only need to use the
most popular $10$ keywords as the common keywords of all the
communities, which gives rise to almost the full prediction and
confirmation of keywords for the un-annotated papers.

The results above show that a community of the citation network can
be interpreted by the most popular $10$ keywords and that the
interpretations of communities can be used in predictions and
confirmations in the network. This experiment shows that for each
community, nodes of the same community do share common features,
that is the short list of common keywords, and that the common
features of each of the communities can be used in predicting and
confirming in networks. Our homophyly model predicts that this
property may be universal for many real networks. The homophyly law
here provides a principle for predicting and confirming functions in
networks.

{\bf Discussions}

We found six principles of structures of networks.
Further exploration of new properties of communities and structures of networks
would build a rich structural theory of networks.
The new principles of structures of networks may play an essential role in
new issues of networks such as: games in networks,
stability, robustness and security of networks, predicting in
networks and controlling of networks. The first question left open by our research is to examine the roles of community structures in
new issues and applications of networks mentioned above. Secondly, our theory can be regarded as a local theory of networks, corresponding to this,
we need a global theory of networks. At last, we define the dimension of a network to be the maximal number of
colors of a node among all nodes of the network. In so doing, we know that the homophyly networks all have dimension one. Therefore, our theory is
a liner network theory. Clearly, it is interesting to develop a non-linear (or high dimensional) network theory.

{\bf Methods }

The data of real network in our keywords prediction can be found from the web sites:
\url{http://snap.standford.edu}, or
\url{http://www-personal.umich.edu/~mejn/netdata}.

{\bf Acknowledgements}

All authors are partially supported by the Grand Project ``Network
Algorithms and Digital Information'' of the Institute of Software,
Chinese Academy of Sciences, by an NSFC grant No. 61161130530 and a
973 program grant No. 2014CB340302. The third author is partially supported by a National Key Basic Research Project of China (2011CB302400)
and  the "Strategic Priority Research Program" of the Chinese Academy of
Sciences£¬Grant No. XDA06010701.

{\bf Author Contributions} AL designed the research and wrote the
paper, JL and YP performed the research.

{\bf Additional information }

Competing financial interests: The authors declare they have no
competing financial interests.

\end{document}